# The inequivalent substitution on Ca site of cubic CaTiO$_3$ perovskite for transparent conductive oxides from USPEX and DFT

Yunting Liang[*]

[1] School of Materials Science and Engineering, Zhengzhou University, Zhengzhou 450001, China

[2] State Centre for International Cooperation on Designer Low-carbon & Environmental Materials, Zhengzhou University, 100 Kexue Avenue, Zhengzhou 450001, China

Corresponding author: yunting_liang@gs.zzu.edu.cn

**Abstract:** Based on genetic evolution algorithm, universal structure predictor USPEX assists to explore the most stable structure from Na$^+$ substituting Ca$^{2+}$ under variable concentration, for p type transparent conductive oxide application. The study finds 2 atoms of Na$_{Ca}$ maintain the octahedral packing and cubic phase without the presence of oxygen vacancy, but the electron loss causes in fermi level entering into valence band, realizing p type TCOs character. While others still remain insulator character due to the compensation of oxygen vacancy. Applying Al$^{3+}$ substituting Ca$^{2+}$ achieve the fermi level into conduction band with n-type TCOs, with the symmetry of tetragonal phase. And the transport properties for p- type Ca$_{0.75}$Na$_{0.25}$TiO$_3$ and n-type Ca$_{0.75}$Al$_{0.25}$TiO$_3$ are calculated, the mobility in p- type Ca$_{0.75}$Na$_{0.25}$TiO$_3$ is more superior owing to its curvature edge.

## Introduction

Presently, transparent conductive oxides (TCOs) serve in a wide variety of applications including devices such as screen panels for televisions, sensors, solar cells, and even photovoltaic windows.[1-3] TCOs are materials that are conductive and at the same time exhibit a transparent characteristic that allows light to pass through the material itself. TCOs are a type of very thin film that is usually deposited on different substrates. Substrates used in solar applications are usually made from material with high transparency such as silica glass or fluorine tin oxide (FTO) glass in order to allow more light to penetrate the material to maximize efficiency.[4,5] The characteristics that are important for a TCO to be suitable for solar applications are the transmittance and sheet resistance. The TCO itself must also have a band gap energy of 3.1 eV or greater, and the high band gap allows the TCO to transmit 80% of visible light or more. [6-8]

The current commercially TCO coatings are dominated by Sn-doped indium oxide (ITO), owing to its high transparency and low resistivity. However, the scarcity of indium has led to the material being expensive in terms of raw material cost and it is a toxic material. FTO is another TCO that is widely used as an alternative to ITO and has a transmittance rivalling that of ITO. However, it has been mentioned in other literature that FTO has relatively low electrical conductivity and is much harder to pattern by wet etching compared to ITO. [9] Another candidate AZO (Al-ZnO) suffers from the setback of low material stability. Great efforts are therefore needed to deliver alternative TCO candidates using low-cost materials with large and green resources, so that sustainable alternative materials can be used to meet the ever-increasing need of TCO coatings. Perovskite with the form of $ABO_3$ was discovered by Gustav Rose, with A being an alkaline-earth metal or alkali metal element and B a transition metal. Components B and O constitute $BO_6$ octahedron backbone, and A fills in the interspace of octahedrons. The electronic structure reveals transition metal (B site) d states locating at the conduction band minimum (CBM) and oxygen 2p states located at the valence band maximum (VBM), which contribute its wide band gap of >3.0 eV.[10,11] This wide band gap of perovskites enables the material to be transparent to wavelengths of >400 nm. This property makes them suitable for use as TCOs in solar harvesting or other photovoltaic applications. However, there have been no perovskite-based TCO that have been implemented into a device yet since more thorough study is required for a perovskite material to be ready to be used in the industry. There have been a few studies on perovskite-structured TCOs and the results were reported on a laboratory scale only, lack of theoretical explanation. [12-17] It is envisaged to design material modelling to efficiently guide the experimental preparation in advance.

Furthermore, these single doping or alloying in modulating TCOs usually company with the presence of vacancy defect for the charge compensation function, which eventually eliminates the doping effect, specially oxygen vacancy for p type doping. Therefore, it is necessary to consider the low dose doping or co-doping for the electric neutrality. Actually, the covalent bond of transition metal and oxygen atoms for perovskite backbone tends to weaken but not to break owing to the covalent electron slight loss from the cation substitution with low chemical valence, vice versa. Based on this point, one concern is raised to doubt the necessity of the charge compensation for single doping or alloying theoretically.

Here we adopt an outstanding and effective structure searching tool-USPEX[18] to identify the presence possibility of oxygen vacancy for scarce p- type perovskite based-TCOs, and investigate the substitution site, which is usually fixed ahead in traditional structure relaxation. Generally speaking, the low valent cation substitution brings about insufficient electron, which is bound to transform the symmetry of origin crystal structure. So USPEX structure searching should be an evident solution for the presence of oxygen vacancy. We employ the most abundant $CaTiO_3$ with 3.5 eV as the parent, and it is an operable solution on inequivalent substitution on Ca site to shift the fermi level with realizing the metallic property, in contrast with Ti and O sites contributing the bandgap. Therefore, Na element is chosen as the low valent cation to obtain p- type TCOs, which poses the close Shannon radii (1.39 Å) with Ca (1.34 Å), and the bigger ionization ability. The variation of Na concentration is used to investigate the watershed of the presence of oxygen vacancy. This work is purposed to provide a fundamental perspective in guiding inequivalent doping or alloying on oxide perovskites to modulate their TCOs application. Based on USPEX tool, the work reveals the substitution site without destroying the octahedral packing, according to Na variable concentration. Furthermore, Al and Ag substituting Ca verify the repeatability of this method, and achieve n-type TCOs in Al substituted system.

**Methods**

Based on the evolutionary optimization algorithm, crystal predictor grogram USPEX was used to investigate the possible stable structure after inequivalent substitution, which has been continuously developed by A. R. Oganov etc. [19,20] Under the purpose of the stable doped structure from the fixed composition searching, atom types and numbers of each species were provided for evolutionary calculation. The 2×2×2 supercell were set to 40 atoms for the convenience of electronic structure calculation, and the atom numbers of Na substituting Ca were respectively 0, 1, 2, 4, 6 and 8 for 6 groups of fixed composition search, where even number was considered for doping symmetry. The lattice parameter of cubic $CaTiO_3$ was taken as the reference of initial value. In the first generation, an initial population of 200 structures was randomly produced by using the space group symmetry, after that, each subsequent generation obtained 60 structures by applying the variation operators- 50% heredity, 20% random, 20% soft mutation (creating a new structure by large atomic displacement along the eigenvectors of the softest phonon modes) and 10% lattice mutation (a new lattice was obtained by applying a distortion defined by a symmetric strain matrix on the old lattice). First principles modelling was carried out in the framework of the density functional theory (DFT), using the Vienna Ab initio Simulation Package code (VASP) with a plane-wave basis set described by the projector augmented wave (PAW) method for the ionic potentials.[21] It was well established that functionals from generalized gradient approximation (GGA) underestimate bandgap values for transition metal oxides due to omission of the nonlocal effect, even though they were fundamentally adequate for structure-energy investigation at the ground states (e.g. formation energy, lattice parameters, semi-quantitatively, and magnetism owing spin polarization).[22-26] Here we used the PBE functional[27,28] for efficient structure-energy modelling to determine the stable structures of doped/alloyed materials. For dependable band structures, we adopted the well tested HSE06 functionals.[29-32] All predicted stable structures were fully relaxed and calculated under PBE method. For the electronic structure calculation, we utilized 520eV kinetic energy cutoff for the plane-wave basis set, and apply 0.01 eV/ Å per atom for convergence of residual forces, while Gaussian method with 0.05eV width for Fermi level smearing. K point grit was chosen 6×6×6 for PBE band calculation. The adopted high symmetry paths for cubic $CaTiO_3$ were X(0.5, 0, 0), R(0.5, 0.5, 0.5), M(0.5, 0.5, 0), G(0, 0, 0) and R(0.5, 0.5, 0.5). K mesh was defined with spacing below 0.03 Å for structure-energy calculations. For the accurate electronic structure, HSE06 method was chose for 2Na doped structure with 25% exchange correlation potential.

**Results and discussion**

**USPEX global searching for Na substitution in $CaTiO_3$**

The stable structures in each $Ca_xNa_{1-x}TiO_3$ phase through inequivalent replacement are identified, using the USPEX method for global energy minimization of each chemical configuration. Such a theoretical approach is particularly useful when little information is known about phase structures in a new material system to be formulated, so that potential phase structures can be predicted with associated properties simulated at 0 K. The predicted stable structures with different Na concentration are showed in Fig .1. We find 0Na doped system (pure $CaTiO_3$) shows slight

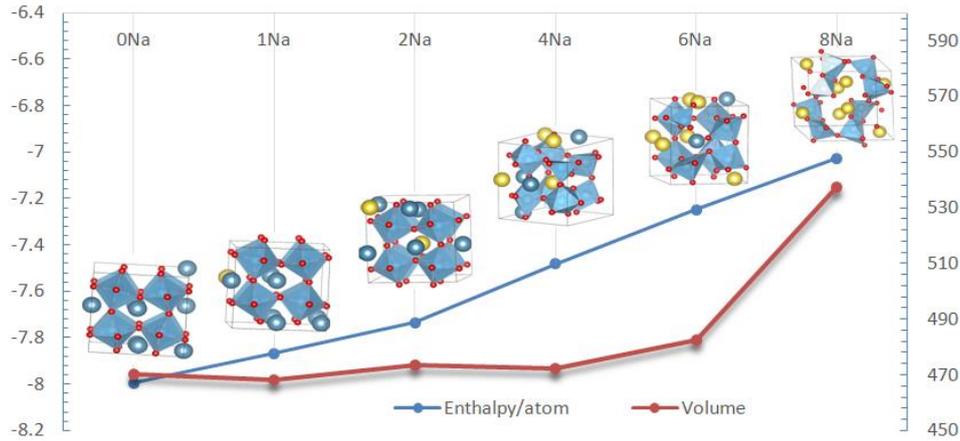

Fig.1. USPEX structure prediction for Na concentration variation in CaTiO$_3$, the y main axis for each atomic enthalpy (eV) and the y secondary axis for the supercell volume (Å$^3$)

tilting in cubic close-packed perovskite structure with TiO$_6$ octahedron constituting structure backbone and alkali metal atoms filling in interspace, and the same situation happens in 1Na and 2Na doped systems without the generation of oxygen vacancy. Meanwhile 4Na and 6Na systems show larger crystal distortions, and these distortions come from TiO$_5$ pyramids, together with oxygen vacancies happened in 4Na system and TiO$_5$ pyramids along with TiO$_4$ tetrahedron appeared in 6Na system. Interestingly, 8Na system forms cyclic structure with the alternant co-existence of the sharing edge of TiO$_6$ octahedron and TiO$_5$ pyramid. Considering Na dopant site, Na atom successfully replaces with Ca position to keep polyhedron stability.

As we know, alkali metal A formed ionic bond with O atom in ABO$_3$ provides electrons to keep the stability of TiO$_6$ octahedron backbone. If the electrons from of alkali metal decrease gradually, TiO$_6$ octahedron structure is certainly bound to this change. Obviously and very convincingly, Na single doping without charge compensation can achieve in 1Na and 2Na doped systems observed in USPEX searched structures after fully searching in 230 group spaces, where electron can still remain original octahedral structure, so this phenomenon proves that charge compensation is lack of necessity for low dose doping. But the occurrence of oxygen vacancy happens in 4Na and 6Na doped systems promotes TiO$_6$ octahedron backbone transforming to TiO$_5$ pyramids, together with the bonding electron loss. This behavior suggests that when bonding charge is not enough to remain the origin structure, the doped systems will automatically create oxygen vacancy, even when we deliberately take no account of considering the charge compensation in initial structure design. When the bonding charge decreases in further, 8Na doped system eventually forms new structure to keep the lowest energy and the most stable structure.

From the trend of thermodynamics results, with the increasing Na content, the enthalpy per atom slightly increases with 0.1eV from 0Na to 2Na doped structure, but starting from 4Na structure increases with 0.2 eV, which predicts more unstable structure. Considering the structure volume and the ionic radius of Na 0.05 Å larger than Ca, doped structures at least maintain close packing to make volume slightly increasing from 0Na to 6Na structures, even along with presence of oxygen vacancy in 4Na and 6Na situations. Undoubtedly 8Na doped system obtains atoms loose packing in new cyclic structure. USPEX prediction provides a better perspective and solution to research doped structures and the occurrence mechanism of oxygen vacancy from chemical bonding.

**Lattice parameters for Na substituted structures**

Lattice parameters of searched stable structures are listed in Table 1. And they show that cubic 0Na doped system obtains smaller lattice than experiment lattice (8.04 Å) with $P\bar{M}3M\_225$ symmetry, and cubic 2Na system shows 0.2% lattice expansion compared with 0Na lattice, because of the ionic radius of Na 0.05 Å larger than Ca. Meanwhile 0.6% c axis shrink in 1Na system makes 1Na lattice pseudo-cubic phase. Undoubtedly the generation of oxygen vacancy for charge compensation in 4Na and 6Na doped systems accelerates larger lattice distortion with elongation along c axis, even though most TiO polyhedrons still remain octahedron condition. Due to the decreasing bonding electron, the symmetry of 8Na lattice is broken drastically.

Table 1 Lattice parameters (Å/ °) and enthalpy for stable and metastable structures (eV) for USPEX searched structures.

| Parameter | 0Na | 1Na | 2Na | 4Na | 6Na | 8Na |
|---|---|---|---|---|---|---|
| a | 7.78 | 7.79 | 7.79 | 7.54 | 7.54 | 6.73 |
| b | 7.78 | 7.79 | 7.79 | 7.74 | 7.65 | 9.34 |
| c | 7.78 | 7.74 | 7.79 | 8.10 | 8.37 | 9.00 |
| α | 90 | 90 | 90 | 90 | 90.20 | 90.70 |
| β | 90 | 90 | 90 | 90 | 89.82 | 92.35 |
| γ | 90 | 90 | 90 | 90 | 89.98 | 71.93 |
| Space group | $P\bar{M}3M\_225$ | P4/MMM_123 | $P\bar{M}3M\_225$ | PMN21_31 | P1_1 | P1_1 |

**Electronic structure calculation for Na substitution**

USPEX searched structures of different Na content are fully relaxed and used to calculate electronic structures from PBE method. The spin-polarized band structures are listed in Fig. 2. Pure CaTiO$_3$ (0Na) shows p type insulator character with Fermi level close to VBM, which is quite common in perovskite materials. With the decreasing electrons in doped systems, the fermi level shifts downwards in valence bands in 1Na and 2Na systems companying with suitable band gap and degrees of magnetism, so 1Na and 2Na systems can be the better candidates for direct p type TCOs. We note that 2Na doped system possess the sharper curve of valence band maximum, which is beneficial to the hole transportation.

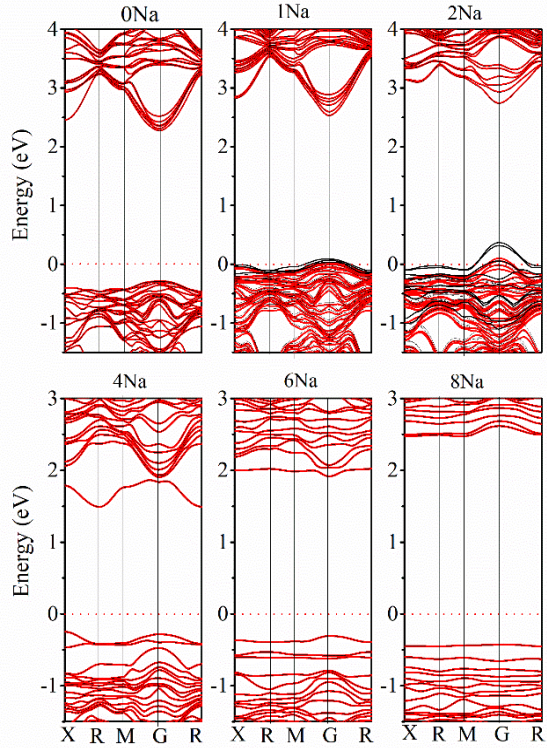

Fig. 2. Spin-polarized band structures from PBE calculation, red lines for spin-up bands and black lines for spin-down bands.

Meanwhile, due to the charge compensation of oxygen vacancy against the electron loss from Na$^+$ substituting for Ca$^{2+}$ in 4Na and 6Na systems, the fermi levels are pulled upwards to form insulator character without the magnetism after the charge balance. With the decreasing bonding electrons, the localization of electron from O 2p orbital (VBM) and Ti 3d orbital (CBM) increases in principle, which results in the flat energy band in 4Na, 6Na doped systems, especially in 8Na system.

2Na predicted structure is further calculated by HSE06 method for more accurate electronic

structure. And the spin polarized band structure and total and partial DOS are showed in Fig.3. From the band structure of 2Na doped $CaTiO_3$ in Fig.3 (a), the spin-down bands dominantly contribute to form the effective band gap with 2.66eV direct transport path in G-G point. To verify the validity of band gap, 0Na structure is treated as the reference with 3.32eV direct band gap obtained by HSE06 method, which is close 0.5eV lower than 3.8 eV experimentally reported value of $CaTiO_3$.[33] After the calibration for HSE06 gap, we can conclude 2Na doped system meet the TCO requirement with 3.16eV gap, Na doping indeed bring oxide semiconductor to p- type transparent conductive oxide without the presence of the compensating charge.

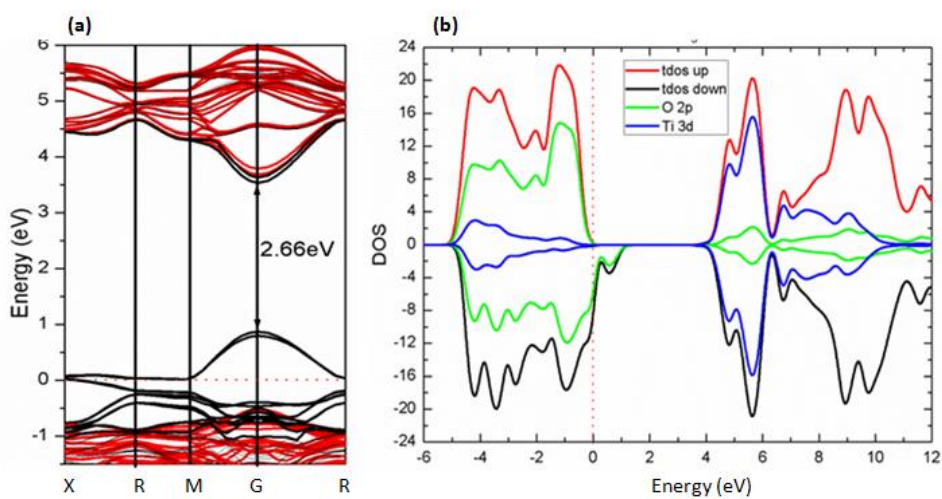

Fig. 3. Spin-polarized band structure (a) and TDOS and PDOS (b) for 2Na doped system from HSE06 calculation.

Further discussion in Fig. 3 (b) shows that Ti 3d energy level contributes to the conductive band maximum and O 2p energy level dominates the valence band minimum, and the majority of PDOS (spin-up) is compensated basically by the minority of PDOS (spin-down) except the range of 0.0 eV ~2.0 eV PDOS. The orbital energy level of Na dopant belongs to the deep energy level, without impacting the forbidden gap, which has no chance showed in Fig.4 (b). But $Na^+$ substituting $Ca^{2+}$ causes the unpaired electron happened to O 2p minority energy level shown in 0.0 eV ~2.0 eV PDOS, and this in further results in minority spin polarization with 2.0 B.M. magnetic moment. Generally, the relationship between the magnetic moment ($\mu_B$) and the unpaired electron amount (n) is calculated by the following formula: $\mu_B =SQRT (n (n+2))$. So 2Na doped system poses one unpaired electron from O 2p orbital contribution.

**Bader charge analysis for Na substitution**

We confirm that the valence electron configurations of Ca, Ti, O and Na are respectively $3s^23p^64s^2$, $3s^23p^63d^24s^2$, $2s^22p^4$, $2p^63s^1$ for Bader charge analysis.[34] Generally, the chemical valence of cation equals to its participating valence electrons minus the final average possessed electrons and the valence of anion equals to the final average possessed electrons minus its participating valence electrons. So electron gain-loss conditions for all considered systems are listed in Table 2. As we know, A-O bond shows ionic bond character and B-O bond holds covalent bond character in $ABO_3$ basic perovskite model. Extending to the $CaTiO_3$ case, Ca atoms provide electrons to keep the $TiO_6$ octahedral stability. Furthermore, we find that Ca and Na cations almost remain the same value with approximate chemical valence $Ca^{2+}$ and $Na^+$. Because of $TiO_6$ octahedron gaining one less electron from alkali metal substitution, from 2Na to 8Na systems, Ti atoms lose more electrons and O atoms gain less electrons to make sure the octahedron stability. When gained electrons of O atoms are less than 1.19, oxygen vacancy happens in 4Na and 6Na systems, and the existing electrons in oxygen atoms become more localized to keep structure stable, which can be verified by bands' curve. Ti and O atoms incline to share electrons in Ti-O covalent bond, so calculated chemical valences are different from traditional valences.

Table 2 Electronic charges of each atomic species in considered systems by Bader charge analysis. The positive (negative) sign represents the gained (lost) electrons.

| Element | 0Na | 1Na | 2Na | 4Na | 6Na | 8Na |
| --- | --- | --- | --- | --- | --- | --- |
| Ca | -1.60 | -1.60 | -1.60 | -1.60 | -1.61 | -- |
| Ti | -2.07 | -2.11 | -2.14 | -2.19 | -2.14 | -2.10 |
| O | +1.22 | +1.21 | +1.19 | +1.14 | +1.07 | +1.00 |
| Na | -- | -0.89 | -0.89 | -0.88 | -0.88 | -0.88 |

**$Ag^+$ and $Al^{3+}$ substituting $Ca^{2+}$ in $CaTiO_3$**

We further investigate the application of this substitution method on other elements with covalent property to modulate n- or p- type TCO, companying with the assumption of $Ag^+$ substituting $Ca^{2+}$ for p- type and $Al^{3+}$ substituting $Ca^{2+}$ for n- type. The stable structures from USPEX global searching are shown from three crystal planes for the clear observation of octahedron distortion. In Fig. 4 (a), Ag atoms take the sites of two face centers and four edges, and they indeed replace the Ca positions without the influence on Ti-O covalent bonding in octahedron cage, in spite that Ag atoms tend to form covalent bond with oxygen. We observe the crystal distortion after Ag substitution from different crystal planes, ab plane shows the regular arrangement of $TiO_6$ octahedrons with a tiny slight deformation, which keeps almost cubic phase character. However, bc and ac planes display a larger distortion, owing to oxygen atoms tending towards Ag atoms. This tendency results in 150° Ti-O-Ti bond angle, with obviously deviating from 180° in cubic phase. Lattice parameters in Table 3 reveal the character of tetragonal phase with I4/MMM(139) space group, with 0.082Å elongation along c axis.

From Al substituted crystal structure in Fig. 4 (b), Al atoms successfully occupy the Ca sites, with the same locations with Ag atoms. ab plane exhibits the opposite tilt angle of $TiO_6$ octahedrons, the front tilt angle is about 130°, and the back angle is 150°, inversely, but bc and ac planes shows almost the same situation. Al substitution causes a larger distortion than Ag substituted structure, predicting the strong bonding inclination of oxygen atoms towards Al atoms. This distortion results in 0.159 Å reduction along c axis with 0.98 tetragonality character, as shown Table 3.

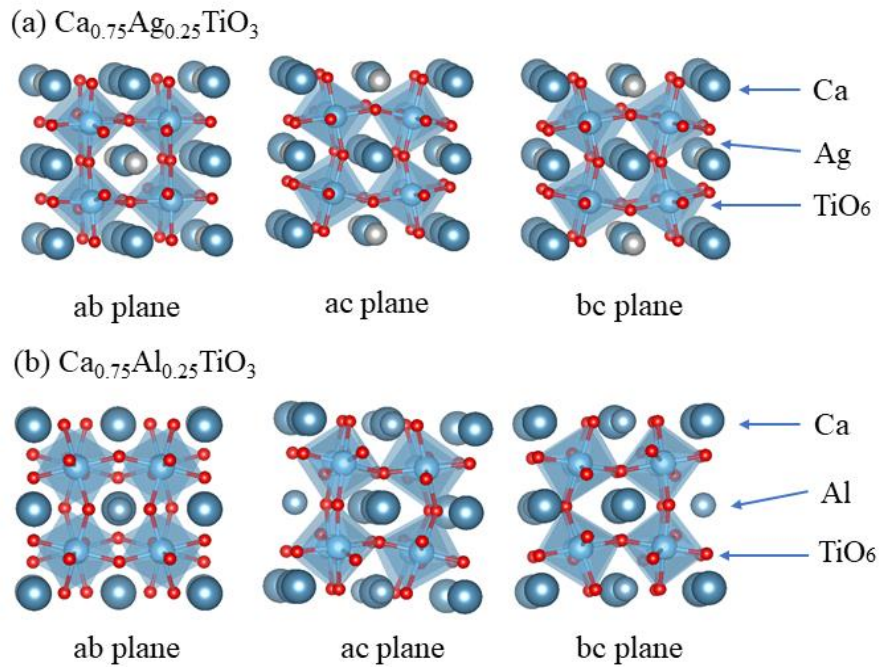

Fig.4. Crystal structures of Ag and Al substituting Ca

Table 3 Lattice constants (Å, °) of Ag and Al doped systems

| | a | b | c | c/a | α | β | γ | Symmetry | Phase |
|---|---|---|---|---|---|---|---|---|---|
| $Ca_{0.75}Ag_{0.25}TiO_3$ | 7.676 | 7.676 | 7.757 | 1.01 | 90 | 90 | 90 | I4/MMM(139) | tetragonal |
| $Ca_{0.75}Al_{0.25}TiO_3$ | 7.684 | 7.684 | 7.525 | 0.98 | 90 | 90 | 90 | P42/NMC(137) | tetragonal |

Electronic structures are calculated to obtain their band structures and density of states to investigate their electronic properties. In band structure of $Ca_{0.75}Ag_{0.25}TiO_3$ in Fig.5 (a), the VBM locates at about -1.0 eV from the contribution of O 2p states, and the CBM composed of Ti 3d states at near 1.0 eV, combined with TDOS and PDOS shown in Fig.5 (b). These impurity bands occurred in forbidden gap come from the hybridization of O 2p and Ag 4d states, playing the dominant role of reducing the bandgap. The direct bandgap shrinks into 0.551eV, while the fermi level appears at the middle of the forbidden gap, without showing the expected p- type property. Furthermore, the VBM sites at -2.75eV and the CBM at -0.2eV, from the PDOS of $Ca_{0.75}Al_{0.25}TiO_3$, Al 3p states locate at deep energy level, without contributing the gap reduction. Besides, the fermi level enters into valence bands, Al substituted system gains 2.414eV direct bandgap with the calibration of bandgap from above discussion in 2Na system, realizing its n- type TCO character.

The Bader charge of Ag and Al cations in their substituted systems are +0.884 and +2.423, respectively, in Table 4. These data reveal the ionic bond nature of Ag and Al cations, in contrast with the calculated chemical valence of Ti and O ions deviating their nominal valence, due to their sharing electrons rather than totally loss. Although Ag and Al cations tend to oxygen anions as shown in crystal structures, the ionic bonding takes the leading role in their bonding ability.

Table 4 Bader analysis of gain and loss of each element in Ag and Al doped systems

| | Ca | O | Ti | A' |
|---|---|---|---|---|
| $Ca_{0.75}Al_{0.25}TiO_3$ | +1.587 | -1.115 | +1.932 | +0.884 |
| $Ca_{0.75}Ag_{0.25}TiO_3$ | +1.551 | -1.218 | +1.885 | +2.423 |

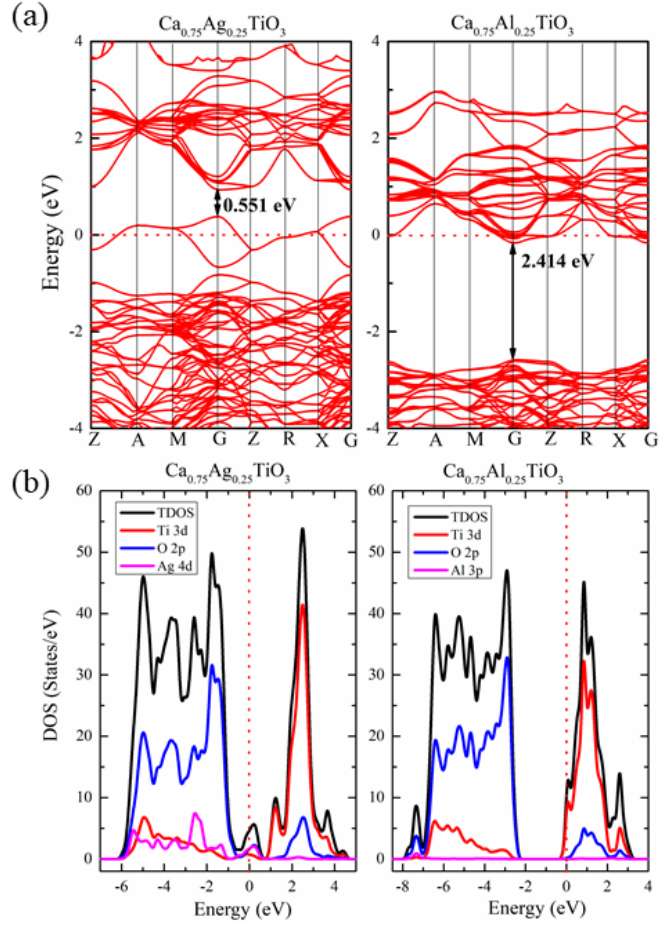

Fig.5. Band structures (a) and TDOS and PDOS (b) for Ag and Al doped systems.

**Transport properties**

Carrier mass and associated mobility are important for p- type $Ca_{0.75}Na_{0.25}TiO_3$ and n-type $Ca_{0.75}Al_{0.25}TiO_3$ for TCOs application. Here a simple approach based on the deformation potential theory is used to examine their transport properties,[35]

$$\mu = \frac{(8\pi)^{1/2}\hbar^4 e c_{ii}}{3(m^*)^{5/2}(k_B T)^{3/2} E_1^2} \qquad (1)$$

where $\hbar$ is reduced Planck constant; e for electronic charge; $c_{ii} = V_0(\frac{\partial^2 E}{\partial V^2})$ for the elastic constant, $V_0$ is the lattice volume; $m^*$ for the carrier effective mass (inversely proportional to the band curvature at CBM or VBM for electrons and holes, respectively); $k_B$ for the Boltzmann constant; T for temperature (300K). $E_1 = \frac{\Delta E_{BM}}{\Delta V/V_0}$ for the deformation potential, the energy shifting of the band edge with respect to the volume variation ($\Delta V$) along the original lattice ($V_0$).

Bulk modulus (B-eV/Å$^3$), deformation potential (DP-eV), effective masses and mobilities ($\mu$-cm$^2$V$^{-1}$s$^{-1}$) for p- type $Ca_{0.75}Na_{0.25}TiO_3$ and n-type $Ca_{0.75}Al_{0.25}TiO_3$ are summarized in Table 5. For Ca$_{0.75}$Na$_{0.25}$TiO$_3$, the evident lifting of the VBM with reference to the case of the host material indicates the tendency for p type characteristics. Na doping leads to VBM to appear at the G point, with the curvature from M point towards the G point direction being much more pronounced than that towards the R point. This causes the effective electron mass is smaller along the M-G direction than the mass along G-R direction, in the reciprocal space. The maximum hole mobility in the M-G direction is almost 300 times of that in silicon. Therefore, Ca$_{0.75}$Na$_{0.25}$TiO$_3$ is to be highly attractive as a TCO based on sustainable materials resources, so that resource crisis due to the depletion of indium resource for ITO can be addressed with a promising alternative. The CBM for Ca$_{0.75}$Al$_{0.25}$TiO$_3$ are located at the G point, which shows more remarkable curvature along the M-G direction than G-Z direction in the reciprocal space, Fig. 6. Radical shift of CBM down below the Fermi energy is induced owing to replace 2/8 of the Ca species in the 2 × 2 × 2 supercell (40 atoms), making the semiconductor into degenerated metallic phases. The effective electron masses are 0.11 along the M-G direction and 0.25 along G-Z, in the reciprocal space. Compared with the effective masses inCa$_{0.75}$Na$_{0.25}$TiO$_3$, the electron mobility in Ca$_{0.75}$Al$_{0.25}$TiO$_3$ are inferior to p- type situation.

Table 5 Bulk modulus (B-eV/Å$^3$), deformation potentital (DP-eV), effective masses and mobilities (μ-cm$^2$V$^{-1}$s$^{-1}$) for p- type Ca$_{0.75}$Na$_{0.25}$TiO$_3$ and n-type Ca$_{0.75}$Al$_{0.25}$TiO$_3$ along with the corresponding orientation.

|  | B | DP | m*(m$_0$) |  | M |  |
|---|---|---|---|---|---|---|
| Ca$_{0.75}$Na$_{0.25}$TiO$_3$ | 0.122 | 11.219 | 0.014 M→G | 0.065 G→R | 116920 M→G | 26274.84 G→R |
| Ca$_{0.75}$Al$_{0.25}$TiO$_3$ | 0.00121 | 6.804 | 0.112 M→G | 0.254 G→Z | 179.731 M→G | 23.178 G→Z |

## Conclusion

We had performed USPEX structure prediction for the lowest enthalpy structures of Na single doped CaTiO$_3$ for p type transparent conductive oxide application. We found that perovskite structure tolerated the bonding electron loss of octahedral backbone 1Na and 2Na doped structures, without the compensation of oxygen vacancy. But with the increasing loss of the bonding electron, the octahedral structure was damaged and the presence of oxygen vacancy happened in 4Na, 6Na and 8Na doped systems. From lattice parameter analysis, 1Na structure obtained pseudo-cubic phase and 2Na structure showed cubic phase. From the electronic structure analysis of all considered systems, 1Na and 2Na doped systems showed metallic character and met the requirement of p type TCO, while others still remained insulator character due to the compensation of oxygen vacancy. Bader charge analysis revealed that with the increasing Na content, the gained electron amount of oxygen anion decreases, which predicted that $TiO_6$ octahedral backbone no longer existed, instead of the generation of $TiO_5$ pyramid. We further investigated the application of this substitution method on other elements (Al and Ag) with covalent property to modulate n- or p- type TCO. Due to Al and Ag atoms inclining to form covalent bond with oxygen, their crystal structures deviated from cubic phase into tetragonal phase. We calculated their transport properties for p- type $Ca_{0.75}Na_{0.25}TiO_3$ and n-type $Ca_{0.75}Al_{0.25}TiO_3$, and the mobility in p- type $Ca_{0.75}Na_{0.25}TiO_3$ is more superior from its curvature edge.

**Notes**

The authors declare no competing financial interest.

**Acknowledgements**